\documentclass[a4paper,twoside,notoc,11pt]{JHEP3}

\usepackage{epsfig,multicol}

\newcommand\fverb{\setbox\pippobox=\hbox\bgroup\verb}
\newcommand\fverbdo{\egroup\medskip\noindent%
			\fbox{\unhbox\pippobox}\ }
\newcommand\fverbit{\egroup\item[\fbox{\unhbox\pippobox}]}
\newbox\pippobox

\newcommand{\nn}{\nonumber}
\newcommand{\beq} {\begin{equation}}
\newcommand{\eeq} {\end{equation}}
\newcommand{\beqa} {\begin{eqnarray}}
\newcommand{\eeqa} {\end{eqnarray}}

\newcommand{\ie}{{\it i.e.}}
\newcommand{\eg}{{\it e.g.}}

\newcommand{\as}{\alpha_s}

\newcommand{\ieps}{i\varepsilon}
\newcommand{\order}[1]{${\cal O}\left(#1 \right)$}

\newcommand{\eq}[1]{(\ref{#1})}

\newcommand{\ket}[1]{\vert{#1}\rangle}
\newcommand{\bra}[1]{\langle{#1}\vert}
\newcommand{\ave}[1]{\langle{#1}\rangle}

\newcommand{\pvec}{\vec p}
\newcommand{\bpsi}{\ol\psi}
\newcommand{\bzet}{\ol\zeta}
\newcommand{\bchi}{\ol\chi}
\newcommand{\bet}{\ol\eta}

\newcommand{\half}{{\scriptstyle \frac{1}{2}}}
\newcommand{\halft}{{\textstyle \frac{1}{2}}}

\newcommand{\ol}{\overline}

\newcommand{\Slash}[1]{ \parbox[b]{0.6em}{$#1$} \hspace{-0.55em}
                            \parbox[b]{0.55em}{ \raisebox{-0.2ex}{$/$}}}

\title{Perturbative QCD with Quark and Gluon Condensates\thanks{This paper is published only in the hep-ph archive. Comments are welcome.}}

\author{Paul Hoyer\thanks{On leave from the Department of Physics,
University of Helsinki, Finland. Research supported in part by the European Commission under contract HPRN-CT-2000-00130.}\\
	Nordita, Blegdamsvej 17, DK-2100 Copenhagen, Denmark\\
	E-mail: \email{hoyer@nordita.dk}}

\preprint{NORDITA-2002-19 HE\\ \hepph{0203236}}

\abstract{QCD perturbation theory using quark and gluon propagators that have an extra term at zero four-momentum is equivalent to standard PQCD in the presence of massless quark and gluon pairs in the perturbative vacuum. We verify at low orders in perturbation theory that $\ave{\ol{\psi}\psi} \neq 0$ and that quarks acquire a constituent mass. Thus chiral symmetry is spontaneously broken, and the conservation of the axial vector current ensures a massless pion. The gluon pairs generate an $\ave{F_{\mu\nu}F^{\mu\nu}}$ expectation value and a tachyonic gluon mass. Lorentz and gauge symmetry is preserved, and the short-distance structure of PQCD is unaffected. The modified perturbative expansion thus contains relevant elements of QCD at both short and long distances.}

\begin{document} 

\section{Perturbing in the presence of parton pairs}

We wish to call attention to the interesting properties of a Perturbative QCD (PQCD) expansion where the free, massless quark and gluon propagators have an extra term at vanishing four-momentum. In Feynman gauge,
\beqa
S_q^{AB}(p) &=& \delta^{AB}\left[\frac{i\Slash{p}}{p^2+\ieps}+C_q (2\pi)^4
\delta^4(p)\right]
\label{qpropmod}\\
D_g^{ab,\mu\nu}(p) &=& -g^{\mu\nu} \delta^{ab}\left[\frac{i}{p^2+\ieps}+C_g
(2\pi)^4 \delta^4(p) \right] \label{gpropmod}
\eeqa
A corresponding modification of the ghost propagator is irrelevant since
the ghost-gluon vertex is proportional to the ghost momentum.

The gluon propagator modification \eq{gpropmod} was proposed in
\cite{cpm}, motivated by a Lorentz and gauge invariant formulation of
QCD vacuum effects. In later work it was shown that such a modification would arise if gluon pairs are present in the asymptotic wave function \cite{rc}, and some of its phenomenological consequences were studied \cite{cr}. Independently, it was noted in \cite{ph1} that perturbative expansions using different $\ieps$ prescriptions at the poles of the free gluon propagator are equivalent to standard expansions with gluons added to the in- and out-states. This was inspired by earlier studies \cite{ph2} of the relation between modifications of the on-shell propagators and boundary conditions.

It should be emphasized that a perturbative expansion using propagators such as \eq{qpropmod} and \eq{gpropmod} is formally as justified as the standard expansion. Standard perturbation theory uses empty, perturbative vacua as in- and out-states at asymptotic times ($t=\pm\infty$). In the infinite time separating the empty states from the scattering event they will relax to the true QCD ground state (assuming the overlap to be non-vanishing). At finite orders of PQCD one sees, however, only the beginning of this relaxation. Thus if the true ground state differs essentially from the perturbative vacuum, \eg, by containing quark and gluon condensates \cite{con}, the perturbative expansion will not capture this physics. This is a likely reason for the failure of standard PQCD at long distances, where soft condensate partons have important effects.

It is thus natural to consider PQCD with asymptotic states that contain (free) quark and gluon pairs. Assuming again an overlap of those states with the true ground state the resulting expansion will be {\em a priori} as justified as standard PQCD. The addition of quark and gluon pairs to the asymptotic states implies no change in the QCD lagrangian nor in the short distance properties of the theory (including renormalizability). Such asymptotic states may on the other hand give rise to non-vanishing quark and gluon condensates, `constituent' parton masses and a spontaneous breaking of chiral symmetry.

There are phenomenological indications that perturbative methods are relevant for long distance QCD physics. Constituent quarks are effective degrees of freedom in hadrons. The transition from short distance (quark and gluon) to long-distance (hadron) physics is smooth \cite{ph1,yd}. This suggests that the strong coupling freezes, $\as(Q^2 = 0) \simeq 0.5$, and that confinement and chiral symmetry breaking is due to quark and gluon `condensates' in the QCD vacuum \cite{con}, rather than to a strong coupling regime.

The abstract contains a detailed study of how gaussian wave functions of fermion and scalar fields at asymptotic times $t=\pm T \to \pm\infty$ imply free propagators of the form \eq{qpropmod} and \eq{gpropmod}. The coefficient of the zero-momentum contribution is determined by the gaussian, and vanishes when the wave function is that of the ground (empty) state. In general the gaussian may be expanded to reveal an indeterminate number of particle pairs.

\section{Vacuum expectation values and constituent masses}

The consequences of the modifications \eq{qpropmod} and \eq{gpropmod} to the quark and gluon propagators are profound. Due to the infinite number of pairs the modified expansion is non-perturbatively related to standard PQCD.

We find that the propagator modifications give rise to non-vanishing $\ave{\ol{\psi}\psi}$ and $\ave{F_{\mu\nu}F^{\mu\nu}}$ expectation values. Thus chiral symmetry is spontaneously broken. Quarks gain a constituent mass whereas gluons become tachyonic due to interactions with the condensates. We verify (at lowest non-trivial order) that the altered boundary conditions do not affect the conservation of the vector and axial currents, $\partial_\mu J^\mu = \partial_\mu J_5^\mu = 0$, which follows from the symmetries of the lagrangian. Thus the pion is massless, despite the masses of its constituent quarks.

\subsection{Quark and gluon condensates}

The quark propagator \eq{qpropmod} leads to spontaneous chiral symmetry breaking already at zeroth order in perturbation theory,
\beq \label{qcond}
\ave{\ol{\psi}(x)\psi(x)} = \sum_{\alpha,A} \frac{-\delta}{\delta\zeta_\alpha^A(x)} \frac{\delta}{\delta\ol\zeta_\alpha^A(x)} \exp\bigg[\int\frac{d^4p}{(2\pi)^4} \ol\zeta^A(p) S_q^{AB} \zeta^B(p) \bigg] = -4NC_q
\eeq
where $N$ is the number of colors. In effect, the divergently large number $\propto 1/m$ of zero-momentum quark pairs implied by Eq. \eq{zcontf} compensates the factor $m$ in the numerator of the propagator, giving a finite result in the $m\to 0$ limit. Thus chiral symmetry is spontaneously broken by the asymptotic state.

In the case of the $\ave{F_{\mu\nu}F^{\mu\nu}}$ expectation value the \order{g^0} contribution vanishes since the free part of $F_{\mu\nu}=\partial_\mu A_\nu -\partial_\nu A_\mu -g f_{abc}A_\mu^b A_\nu^c$ is proportional to the gluon momentum. Furthermore, at \order{g^2} the gluon loop integrals vanish (when dimensionally regularized) due to the absence of a mass or momentum scale. This leaves only the two-loop diagram involving a 4-gluon coupling (Fig. 1a), with the $C_g$ part of the propagator taken in both loops. The result is \cite{cpm,cr}
\beqa \label{gcond}
\ave{F_{\mu\nu}(x)F^{\mu\nu}(x)} &=& g^2 f_{abc} f_{ade} \frac{\delta^4}{\delta J_\mu^b \delta J_\nu^c \delta J^\mu_d \delta J^\nu_e}\exp\bigg[-\frac{1}{2} \int \frac{d^4p}{(2\pi)^4} J(-p) D_g J(p) \bigg] \nn\\
&=& 12g^2 N(N^2-1)C_g^2
\eeqa
The gluon condensate thus arises specifically from the non-abelian 4-gluon interaction.

\EPSFIGURE{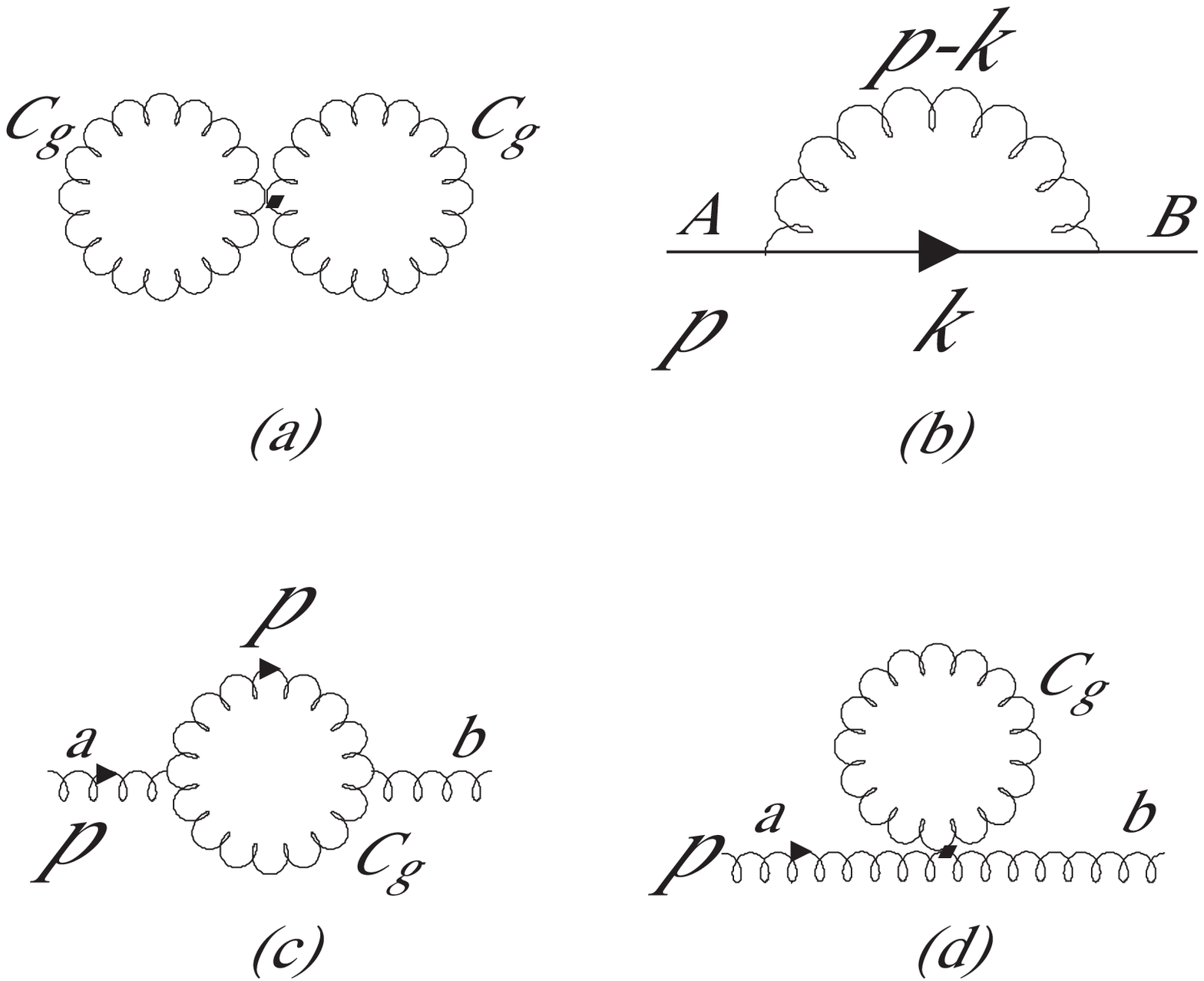,width=12cm}{(a) The only diagram which contributes to $\ave{F_{\mu\nu}F^{\mu\nu}}$ at \order{g^2}. The gluon condensate term $\propto C_g$ in \eq{gpropmod} is taken in both loops. (b) Quark propagator correction. (c,d) Gluon propagator corrections.}

\subsection{Quark constituent mass}

The quark propagator correction $\Sigma(p)$ shown in Fig. 1b gets a contribution from both the quark condensate term $C_q$ in \eq{qpropmod} ($k=0$) and from the gluon propagator modification $C_g$ of \eq{gpropmod} ($k=p$). Only the former generates a proper mass term, making ${\rm Tr} S_q(p) \neq 0$, where (for $p\neq 0$)
\beq \label{qpropcorr}
S_q(p) = \frac{i}{\Slash{p}+\Sigma +\ieps}
\eeq
It is straightforward to write down the quark condensate contribution to the diagram of Fig. 1b,
\beq
\Sigma_q^{AB}(p) = 4g^2 C_F C_q \frac{\delta_{AB}}{p^2+\ieps}
\eeq
where $C_F = (N^2-1)/2N$. As expected, it is suppressed in the short distance $(p^2 \to \infty)$ limit. The pole in the propagator \eq{qpropcorr} is at the `constituent' quark mass $p^2=M_q^2$, with
\beq \label{constmass}
M_q = \Big(4g^2C_F|C_q|\Big)^{1/3}
\eeq
The gluon condensate correction to the quark propagator is similarly \cite{cr},
\beq
\Sigma_g^{AB}(p) = -2g^2 C_F C_g \frac{\Slash{p}}{p^2+\ieps}\delta^{AB}
\eeq
and influences the quark constituent mass $M_q$, which is now is obtained as the solution of the cubic equation,
\beq \label{constmassg}
M_q^3-2g^2C_FC_gM_q = \pm 4g^2C_FC_q
\eeq

\subsection{A tachyonic gluon}

The gluon propagator corrections shown in Figs. 1c,d were evaluated in Ref. \cite{cr}. The sum of the two diagrams has the transverse structure required by gauge invariance. The resummed propagator thus has the form (for $p\neq 0$)
\beq
D_g^{ab,\mu\nu}(p)= i\delta^{ab}\frac{-g^{\mu\nu}+p^\mu p^\nu /p^2}{p^2+\Pi_g +\ieps} - i\delta^{ab}\frac{p^\mu p^\nu}{p^4}
\eeq
where the gluon condensate contribution is
\beq
\Pi_g(p) = 2g^2 N C_g
\eeq
There is no contribution to the gluon propagator from the quark condensate term $C_q$ at \order{g^2}. The insertion of this term in the quark loop correction gives a vanishing trace due to an odd number of  $\gamma$ matrices.

A modification of the ghost propagator at $p=0$ also does not contribute to $\Pi_g$ since the ghost-gluon vertex vanishes at zero ghost momentum. In Ref. \cite{hr}, where asymptotic states of gluons with $p\neq 0$ were considered, the ghost contribution was non-vanishing and essential for gauge invariance.

As was found also in \cite{hr}, the gluon propagator has a pole at negative $p^2$, corresponding to a `tachyonic' gluon mass,
\beq \label{gluemass}
M_g^2 = -2g^2 N C_g
\eeq
provided $C_g > 0$ \cite{rc}. This is likely to have important consequences. A quark or gluon can acquire an absorptive part by emitting a spacelike gluon with $p^2 \simeq M_g^2 <0$. Thus colored partons cannot propagate long distances due to their interactions with the quarks and gluons in the asymptotic state. Any stable eigenmodes of propagation are likely to be color singlets, which do not emit soft gluons.

\EPSFIGURE{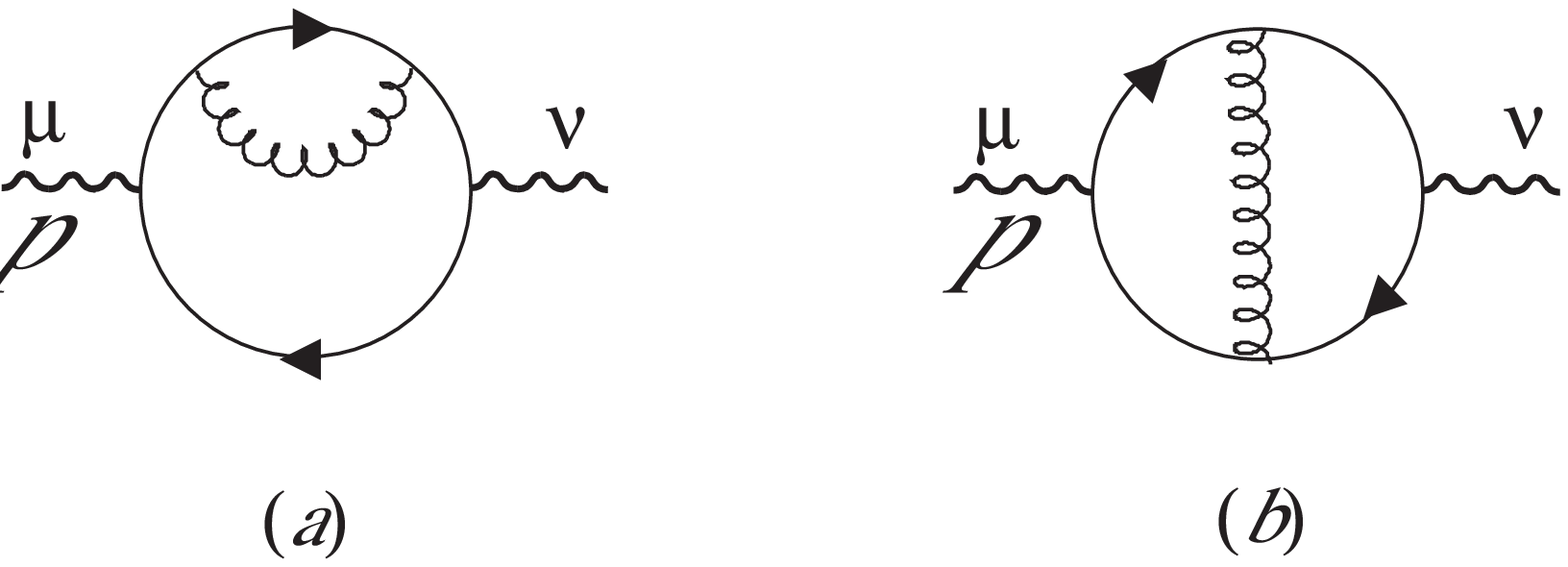,width=12cm}{Corrections to the color singlet (axial) vector current. A third diagram where the gluon couples to the antiquark is not shown.}

\subsection{Conservation of vector and axial vector currents}

The breaking of chiral symmetry \eq{qcond} resulted from our modification of the asymptotic states. The QCD lagrangian is unchanged and remains chirally symmetric. Consequently the symmetry is broken spontaneously and there must be a massless goldstone boson, the pion, which is annihilated by the axial current,
\beq \label{axpion}
\bra{0}J_5^\mu(x)\ket{\pi(p)} = -ip^\mu f_\pi \exp(-ip\cdot x)
\eeq
The symmetry of the lagrangian implies $\partial_\mu J_5^\mu =0$, \ie, $m_\pi = 0$. Current conservation should hold order by order in the modified perturbative expansion. We shall verify it for the diagrams of Fig. 2.

The quark condensate term $C_q$ of \eq{qpropmod} must be used in two of the four quark propagators in Fig. 2 to have an even number of $\gamma$ matrices in the quark loop. Consequently the two-loop integral reduces to a number, and we find, for a vector $(-ie\gamma^\mu)$ coupling of the current,
\beq \label{vecorr}
A_q^{\mu\nu}(p) = 16e^2g^2 (N^2-1) \frac{C_q^2}{p^4}\left(g^{\mu\nu} - \frac{p^\mu p^\nu}{p^2}\right)
\eeq
The transverse structure ensures vector current conservation and gauge invariance. The result for an axial vector current coupling $(-ie\gamma_5\gamma^\mu)$ is the same, up to a sign.

The contribution of the gluon propagator modification \eq{gpropmod} on the gluon line in Fig. 2 reduces the two loop integrals to a single (regular) integral. For a vector coupling we find
\beq
A_g^{\mu\nu}(p) = \frac{e^2g^2}{\pi^2} (N^2-1) C_g\left(g^{\mu\nu} - \frac{p^\mu p^\nu}{p^2}\right)
\eeq
We have again $p_\mu A_g^{\mu\nu}(p) =0$ as required by current conservation. The result is the same for an axial vector current.

\section{Concluding remarks}

The perturbative expansion is a powerful tool in field theory. It is specified by the lagrangian and by the state which one chooses to expand around. The sum is formally independent of this state provided it is taken  at asymptotic times ($t\to\pm\infty$) and has an overlap with the true vacuum. Hence it is interesting to consider the properties of perturbative QCD expansions with non-empty asymptotic states.

A PQCD expansion can be phenomenologically useful provided quarks and gluons are relevant degrees of freedom for the physics being studied. In the short distance regime this is clearly so, and PQCD has been impressively successful. While hadrons in many respects are good degrees of freedom at long distances, there are indications that partons maintain their identity even in this regime. Thus hadrons can be characterized as loosely bound states of constituent quarks, which differ from current quarks only in their mass. It has also been noted that the transition from the short distance to the `confining' regime is smooth \cite{ph1,yd}. This is consistent with a picture where the strong coupling freezes at a perturbative value $\as(Q^2=0) \simeq 0.5$ and the novel long distance effects result from interactions with quark and gluon condensates in the physical ground state.

We previously studied \cite{ph1,hr} the effects of including finite momentum quarks and gluons in the asymptotic states. Such states are not boost invariant and Lorentz invariance is thus implicit, \ie, the convergence of the expansion is frame dependent. This may not be a disadvantage when a choice of frame is anyway implied, as in the calculation of bound states wave functions. However, the possibility of retaining explicit Lorentz invariance by including only massless partons with vanishing momenta \cite{rc,cr} is attractive. This is also similar to the framework of the QCD sum rules \cite{con}, where no momentum is exchanged with the vacuum condensates. 

The modification of the gluon propagator \eq{gpropmod} in Feynman gauge implies a corresponding modification of the ghost propagator. This was shown to give a gauge invariant result in a one-loop calculation of the static potential \cite{hr}. However, since the ghost gluon vertex is proportional to the ghost momentum a modification of the ghost propagator at $p=0$ is irrelevant. The results obtained here and in Ref. \cite{cr} without ghost modifications are indeed consistent with gauge invariance.

In this paper we extended the considerations of Refs. \cite{rc,cr} to quarks and found that chiral symmetry breaks spontaneously. We made non-trivial checks that identities following from symmetries of the lagrangian are respected by the condensate contributions. In particular, we saw how the pion can stay massless despite the finite mass of its constituent quarks. The quark mass contribution (Fig. 2a) is exactly cancelled by the interaction between the quarks in (Fig. 2b).

The fact that the gluon propagator acquires a pole for $p^2 < 0$ (a tachyonic gluon mass) \cite{cr,hr} is likely to have important consequences. It implies that quarks and gluons can propagate only over a limited range in the condensate, as the emission of long wavelength spacelike gluons generates an absorptive part. Since the colored condensate will be more transparent to color singlet bound states it is not farfetched to speculate that hadrons are the only stable eigenmodes of propagation.

\acknowledgments

I am grateful for discussions with St\'ephane Peign\'e and Johan Rathsman.

\bigskip\bigskip
\centerline{APPENDIX}

\appendix

\section{Influence of the asymptotic state on the free propagator}
\label{sec:A}

The generating functional $Z[J]$ of Green functions in (scalar) field theory can be expressed in the form
\beq \label{zint}
Z_{int}[J]= \exp\left[iS_{int}\left(\frac{\delta}{\delta J}\right)\right] Z[J]
\eeq
where $J$ is the source of the field $\phi$, $S_{int}(\phi)$ is the interaction part of the action and the free functional is explicitly given by the free propagator $D(p)$,
\beqa 
Z[J] &=& \int {\cal D}[\phi] \exp\left\{i\int d^4x \left[{\cal L}_0(\phi) + J(x)\phi(x)\right]\right\} \label{zfreex} \\
&=& \exp\left[-\frac{1}{2}\int\frac{d^4p}{(2\pi)^4} J(-p) D(p) J(p) \right]
\label{zfreep}
\eeqa

The boundary condition at $x^0 = \pm\infty$ of the path integral \eq{zfreex} is usually specified implicitly by tilting the $x^0$ integration path to the real axis, giving the Feynman propagator
\beq \label{feynprop}
D_F(p)=\frac{i}{p^2-m^2+\ieps}
\eeq
In the operator formalism this propagator is obtained by requiring the perturbative vacuum to be empty, $a\ket{0}=0$.

We wish to show how the inclusion of parton pairs in the asymptotic state modifies the prescription at the $p^2=m^2$ pole of the free propagator. We consider first the case of a scalar field and then that of a fermion, making use of the results in \cite{ph2}. The generalization to the free gluon and quark propagators of QCD is straightforward. The full QCD generating functional is given by the free one as in Eq. \eq{zint}. The standard Feynman rules of PQCD are thus modified only through the change in the free propagators.

\subsection{Free scalar field}

The space part of the gaussian path integral in \eq{zfreex} is diagonalized in 3-momentum space. The following analysis is thus independent of $\pvec$. For simplicity, and since we shall be interested in zero momentum pairs, we assume $\pvec=0$ and treat only the time/energy dimension explicitly.

To see the influence of the asymptotic states on the free propagator we need to first calculate the path integral over a finite time interval, $-T \leq t \leq T$, and then take $T\to\infty$. We define \eq{zfreex} to be
\beqa \label{zfreet}
Z[J] = \int {\cal D}[\phi(t)] \exp\left\{\frac{-i}{2}\int_{-T}^T dt\, \phi(t)\bigg[\frac{\partial^2}{\partial t^2} \right. &+& m^2 \bigg] \phi(t) + i\int_{-T}^T dt\, J(t)\phi(t) \nn\\
 &-&\left. \frac{C_s}{2}\Big[\phi^2(-T)+\phi^2(T)\Big]\right\}
\eeqa
Here $C_s > 0$ parametrizes the gaussian wave functions at $t=\pm T$. Since the free scalar lagrangian describes a set of harmonic oscillators we expect that the ground state is obtained for $C_s=m$. We find below that precisely this value of $C_s$ gives the Feynman propagator. Other values of $C_s$ correspond to a superposition of wave functions containing particle pairs.

Describing the field through its Fourier components,
\beq \label{fourier}
\phi(t)= \frac{1}{2T}\sum_{p=-\infty}^\infty \phi_p \exp(-i\pi pt/T)
\eeq
implies a periodic boundary condition,
\beq \label{phiT}
\phi(T)=\phi(-T)=\frac{1}{2T}\sum_{p=-\infty}^\infty (-1)^p \phi_p
\eeq
Introducing this relation via a $\delta$-function constraint gives (up to an overall normalization),
\beqa \label{zdiscrp}
Z[J] &=& \int_{-\infty}^\infty d\alpha\,d\phi(T)\,\prod_{p=-\infty}^\infty d\phi_p \exp\bigg\{ \frac{i}{4T} \sum_p \phi_{-p}\left( \frac{\pi^2 p^2}{T^2}-m^2 \right)\phi_p \nn\\
&+& \frac{i}{2T} \sum_p \left[J_{-p}-(-1)^p\alpha\right]\phi_p
-C_s\phi^2(T)+i\alpha\phi(T)\bigg\}
\eeqa
The result of the gaussian integration is
\beqa \label{zdiscres}
Z[J] = \exp\Bigg\{ -\frac{i}{4T}\sum_{p=-\infty}^\infty \frac{J_{-p} J_p} {\frac{\pi^2 p^2}{T^2}-m^2}
-\left[\frac{1}{C_s}+\frac{i}{T}\sum_p \frac{1}{\frac{\pi^2 p^2}{T^2}-m^2}
\right]^{-1} \left[ \frac{1}{2T}\sum_p \frac{(-1)^p J_p}{\frac{\pi^2 p^2}{T^2}-m^2} \right]^2 \Bigg\}\nn\\&&
\eeqa
This is a well-defined expression if $mT \neq \pi p$ for any integer $p$. In order for the $T \to \infty$ limit to be smooth we take
\beq \label{mTconstr}
mT = \left(N+\frac{1}{2}\right)\pi 
\eeq
which ensures that the denominators vanish in the middle of a discrete momentum interval, and then let $N \to \infty$. 

Using
\beq \label{sump}
\sum_{p=-\infty}^\infty \frac{1}{p^2-\alpha^2} 
= \cos(\pi\alpha) \sum_{p=-\infty}^\infty \frac{(-1)^p}{p^2-\alpha^2}
= - \frac{\pi\cos(\pi\alpha)}{\alpha\sin(\pi\alpha)}
\eeq
implies with \eq{mTconstr} that
\beq
\sum_{p=-\infty}^\infty \frac{1}{\frac{\pi^2 p^2}{T^2}-m^2} = 0
\eeq
in Eq. \eq{zdiscres}. In the continuum $(N\to\infty)$ limit the source $J_p$ is slowly varying compared to $(-1)^p$ in the last term of \eq{zdiscres}. Hence that term only contributes for $p\simeq mT/\pi$, where the denominator is rapidly varying. Thus, using again \eq{sump},
\beq \label{polecont}
\lim_{N\to\infty}\frac{1}{2T}\sum_p \frac{(-1)^p J_p}{\frac{\pi^2 p^2}{T^2}-m^2} = \frac{J(m)}{2T}\sum_p \frac{(-1)^p}{\frac{\pi^2 p^2}{T^2}-m^2} = \frac{J(m)}{2m}(-1)^{N+1}
\eeq
where according to \eq{fourier} we indicate the continuum energy $E$ as $J_p = J(E=\pi p/T)$. Similarly we note that the continuum limit of the first term in \eq{zdiscres} is
\beqa
\lim_{N\to\infty} \frac{-i}{4T}\sum_{p=-\infty}^\infty \frac{J_{-p} J_p}{\frac{\pi^2 p^2}{T^2}-m^2}
&=&-\frac{i}{2}{\cal P}\,\int_{-\infty}^\infty \frac{dE}{2\pi} \frac{J(-E)J(E)}{E^2-m^2}\\
 &-&i\frac{J(-m)J(m)}{4T}\sum_{p=-\infty}^\infty \frac{1}{\frac{\pi^2 p^2}{T^2}-m^2}\nn
\eeqa
where ${\cal P}$ denotes the principal value. The second term represents the contribution close to the pole of the propagator where the sum may not be approximated by the integral. This term vanishes due to \eq{mTconstr} and \eq{sump}. Collecting the above results gives
\beqa \label{bosres}
Z[J] &=& \exp\left\{-\frac{i}{2}{\cal P}\,\int_{-\infty}^\infty \frac{dE}{2\pi} \frac{J(-E)J(E)}{E^2-m^2} -\frac{C_s}{4m^2}J^2(m)\right\} \nn\\ &&\\
&=&\exp\left\{-\frac{1}{2}\int_{-\infty}^\infty \frac{dE}{2\pi} J(-E) \left[\frac{i}{E^2-m^2+\ieps} +\frac{1}{2m}\left(\frac{C_s}{m}-1\right) 2\pi\delta(E-m)\right]J(E)\right\}\nn
\eeqa
Note that $J(-m) = J(m)$, since only the combination $J_p+J_{-p}$ enters in the pole contribution \eq{polecont}. This is a consequence of the periodic boundary condition \eq{phiT}. In general $\phi(t)$ is related to $\phi(-t)$ through $\phi_p \to \phi_{-p}$, hence the asymptotic momentum space field $\phi(m)$ is real.

Eq. \eq{bosres} shows that the Feynman propagator is modified for $C_s \neq m$, analogously to the gluon propagator in Eq. \eq{gpropmod}. As we noted above, $C_s=m$ corresponds to imposing the ground state wave function as boundary condition at $t=\pm T \to \pm \infty$. For $C_s \neq m$ the gaussian wave function may be expanded in powers of $(C_s-m)\phi^2(\pm T)$ and interpreted as a state with additional particle pairs.

It is straightforward to extend this analysis to fields $\phi_\lambda(t,\pvec)$ having non-vanishing 3-momenta $\pvec$ and several spin components $\lambda$. If the argument of the gaussian wave function is of the form $\sum_\lambda\phi_\lambda(\pm T,-\pvec) \phi_\lambda(\pm T,\pvec)$ it will represent spin zero pairs with vanishing total 3-momentum. 

An application to the gluon propagator \eq{gpropmod} is more subtle due to gauge invariance. In the limit of vanishing mass $m$ we must let $C_s \to m$ in \eq{bosres}. The sign of $C_g$ in \eq{gpropmod} then depends on whether $C_s$ approaches $m$ from below or above. In the QCD analysis of Ref. \cite{rc} it was found that $C_g > 0$. We use Feynman gauge as in \cite{rc,cr}. Gauge independence was explicitly verified \cite{hr} for Feynman and Coulomb gauge in a one-loop calculation of the static potential with boundary conditions modified at $\vec p \neq 0$. In that case a modification of the ghost propagator analogous to \eq{gpropmod} was essential in Feynman gauge. It is, however, irrelevant in our present ($\pvec=0$) analysis since the ghost-gluon vertex is proportional to the ghost momentum.

\subsection{Spin $\halft$ field}

We take the finite time, free generating functional analogous to \eq{zfreet} for a spin $\half$ field to be
\beqa \label{zferm}
Z[\bzet,\zeta] = \int {\cal D}[\bpsi,\psi] \exp&\bigg\{&\int_{-T}^T dt\,\Big[i\bpsi(t) (i\gamma^0 \partial_t -m)\psi(t) +\bzet\psi + \bpsi\zeta \Big]\nn\\
&-& C_f\Big[ \bpsi(-T)\psi(-T) + \bpsi(T)\psi(T)\Big]\bigg\}
\eeqa
where the fields $\psi,\bpsi$ and sources $\zeta, \bzet$ are Grassmann valued Dirac spinors, and the wave functions at $t=\pm T$ are parametrized by the constant $C_f$. For notational simplicity we do not indicate the 3-momentum (and take $\pvec=0$). The Dirac algebra is diagonalized using fields $\chi_{\sigma,\lambda},\ \bchi_{\sigma,\lambda}$ where $\sigma= +\ (-)$ is an (anti)particle label and $\lambda= \uparrow,\downarrow$ indicates spin component,
\beq \label{psidef}
\psi(t)= \sum_\lambda \Big[u(\lambda)\chi_{+,\lambda}(t)+v(\lambda)\chi_{-,\lambda}(t)\Big]
\eeq
Here $u^\dag(\uparrow)=(1,0,0,0),\ u^\dag(\downarrow)=(0,1,0,0)$, and similarly in terms of the lower components for $v(\lambda)$.

Defining the source $\eta_{\sigma,\lambda}$ correspondingly,
\beq \label{setdef}
\zeta(t)=\gamma^0 \sum_\lambda \Big[u(\lambda)\eta_{+,\lambda}(t) +v(\lambda)\eta_{-,\lambda}(t)\Big]
\eeq
the free fermion functional \eq{zferm} becomes
\beqa \label{zfermdiag}
Z[\bet,\eta] = \int {\cal D}[\bchi,\chi] \exp&\bigg\{& \sum_{\sigma,\lambda} \int_{-T}^T dt\,\Big[i\bchi_{\sigma,\lambda}(t) (i\partial_t - \sigma m)\chi_{\sigma,\lambda}(t) +\bet_{\sigma,\lambda}\chi_{\sigma,\lambda} + \bchi_{\sigma,\lambda}\eta_{\sigma,\lambda} \Big]\nn\\
&-& C_f\sum_{\sigma,\lambda} \sigma\Big[ \bchi_{\sigma,\lambda}(-T)\chi_{\sigma,\lambda}(-T) + \bchi_{\sigma,\lambda}(T)\chi_{\sigma,\lambda}(T)\Big]\bigg\}
\eeqa
In terms of the Fourier components defined as in \eq{fourier} we get,
\beqa \label{zfermp}
Z[\bet,\eta] &=& \int \prod_p d\bchi_p\,d\chi_p \exp\bigg[-\sum_{p,q}\bchi_{p} A_{pq}\chi_{q} +\frac{1}{2T}\sum_p(\bet_{p}\chi_{p} + \bchi_{p}\eta_{p}) \bigg] \nn\\
&=& \det(A)\exp \bigg[\frac{1}{(2T)^2}\sum_{p,q}\bet_{p} A^{-1}_{pq}\eta_{q} \bigg]
\eeqa
where we suppressed (the sum over) $\sigma,\lambda$ in the exponent. The inverse of the matrix
\beq \label{Amatr}
A_{pq}=-\frac{i}{2T}\left(\frac{\pi p}{T}-\sigma m\right)\delta_{pq} + \frac{\sigma C_f}{2T^2}(-1)^{p+q}
\eeq
can be readily guessed from the corresponding result \eq{zdiscres} for scalar fields. Imposing again \eq{mTconstr} and using \eq{sump} we find (for each $\sigma,\lambda$ and ignoring the overall normalization)
\beq
Z[\bet,\eta] = \exp\bigg[\frac{i}{2\pi}\sum_{p=-\infty}^\infty\,\frac{\bet_{p} \eta_{p}}{p-\sigma mT/\pi} +\frac{\sigma C_f}{2\pi^2}\sum_{p,q}(-1)^{p+q} \frac{\bet_{p}\eta_{q}}{(p-\sigma mT/\pi)(q-\sigma mT/\pi)} \bigg]
\eeq
In the continuum $(N \to \infty)$ limit we get, similarly to the scalar case above,
\beqa \label{zcontf}
Z[\bet,\eta] &=& \exp\bigg\{\sum_{\sigma,\lambda}\bigg[i{\cal P}\int_{-\infty}^\infty\frac{dE}{2\pi}\, \frac{\bet_{\sigma,\lambda}(E) \eta_{\sigma,\lambda}(E)}{E- \sigma m} +\frac{\sigma C_f}{2} \bet_{\sigma,\lambda}(\sigma m) \eta_{\sigma,\lambda}(\sigma m) \bigg]\bigg\}\nn\\
 &&\\
&=& \exp\bigg\{\int_{-\infty}^\infty\frac{dE}{2\pi}\, \bzet(E) (E\gamma^0+m)\bigg[{\cal P} \frac{i}{E^2-m^2}+C_f\pi\delta(E^2-m^2)\bigg]\zeta(E) \bigg\} \nn
\eeqa
where $C_f = 1$ gives the Feynman propagator. In the $m\to 0$ limit $\zeta(\pm m) \to \zeta(0)$ and the $E\gamma^0$ term cancels in the sum of the $E=\pm m$ pole contributions. Thus
\beq
Z[\bet,\eta] {\buildrel {m\to 0} \over {=}} \exp\bigg\{ {\cal P}\int_{-\infty}^\infty\frac{dE}{2\pi}\,\bzet(E) \frac{iE\gamma^0}{E^2}\zeta(E) +\halft C_f \bzet(0)\zeta(0)\bigg\}
\eeq
which gives a free fermion propagator of the form \eq{qpropmod}.


\end{document}